\documentclass[10pt]{article}
\usepackage[OE]{express}
\usepackage{siunitx}

\begin{document}
\title{Empirical formulae for hollow-core antiresonant fibers: dispersion and effective mode area}
\author{Md Imran Hasan,\authormark{*} Nail Akhmediev, and Wonkeun Chang}
\address{Optical Sciences Group, Research School of Physics and Engineering, The Australian National University, Acton ACT 2601, Australia}
\email{\authormark{*}imran.hasan@anu.edu.au}

\begin{abstract}
We present empirical formulae that can provide dispersion and average effective area of the fundamental mode in hollow-core antiresonant fibers. The formulae draw on the structural parameters of the fiber, and allow one to obtain the guiding properties over a wide spectral bandwidth, without the need for time consuming numerical simulations. The formulae are validated by comparing their results with those obtained using a finite-element method. We also analyze the effects of changing the number of antiresonant tubes, as well as adding nested elements in the antiresonant tubes on the guiding properties.
\end{abstract}


\section{Introduction}

Hollow-core fibers permit the propagation of light in the central hollow region. This unique feature has potential to open up various applications, including the high-energy and ultrashort pulse delivery \cite{Knight2007}, high speed data transmission \cite{Poletti2013} and particle guidance using laser light \cite{Renn1999}. Moreover, hollow-core fibers can be filled with gas or liquid, which presents an excellent testbed for investigating ultrafast nonlinear light-matter interactions \cite{Travers2011}. The platform offers a wide variety of new applications, such as gas-based lasers \cite{Nampoothiri2010}, in-fiber high-harmonic generation \cite{Heckl2009} and supercontinuum generation \cite{Travers2011,Hasan2016}.

There are two main types of hollow-core fibers that have been developed so far. The first is the hollow-core photonic bandgap fibers, which exhibit very low transmission loss, but have rather limited bandwidth \cite{Poletti2013}. The second is the hollow-core antiresonant fibers (HC-ARFs). Although HC-ARFs have moderately higher transmission loss compared to the hollow-core photonic bandgap fibers, they offer ultrabroadband guidance, high optical-damage threshold and low group-velocity dispersion (GVD) \cite{Belardi2015}. Among the large variety of HC-ARFs, the negative-curvature fibers have recently emerged, featuring exceptional transmission properties \cite{Kolyadin2013,Yu2016,Hasan2017,Poletti2014}. Currently, there is a big motivation among the researchers to study the guiding properties of the negative-curvature HC-ARFs, both experimentally \cite{Kolyadin2013,Yu2016} and theoretically \cite{Finger2014,Vincetti2016}.

One way to obtain the guiding properties of HC-ARFs accurately is to use numerical approaches. However, the geometrical complexity of HC-ARFs often leads to large computer resource requirements. On the other hand, a capillary approximation can be used conveniently to get the guiding properties \cite{Finger2014,Marcatili1964}. In this case, a careful formulation of the approximations is crucial for obtaining accurate results. Marcatili and Schmelzer derived analytical formulae for dispersion and confinement loss of hollow dielectric waveguides, which work well for simple capillary fibers \cite{Marcatili1964}. Travers et al.~utilized this model and employed a different core approximation for use in the dispersion of kagome-lattice fibers \cite{Travers2011}. Later, Finger et al.~proposed an empirical formula for obtaining more accurate GVD of kagome-lattice fibers \cite{Finger2014}. More recently, Vincetti presented an empirical formula for the confinement loss of single-element negative-curvature fibers \cite{Vincetti2016}.

In this work, we present empirical formulae describing the guiding properties of HC-ARFs. Namely, our model introduces a GVD formula that is accurate over a wider spectral range. Additionally, we introduce a formula for obtaining the average effective mode area of HC-ARFs.

\section{Fiber structure}

\begin{figure}[h] 
\centering
\includegraphics{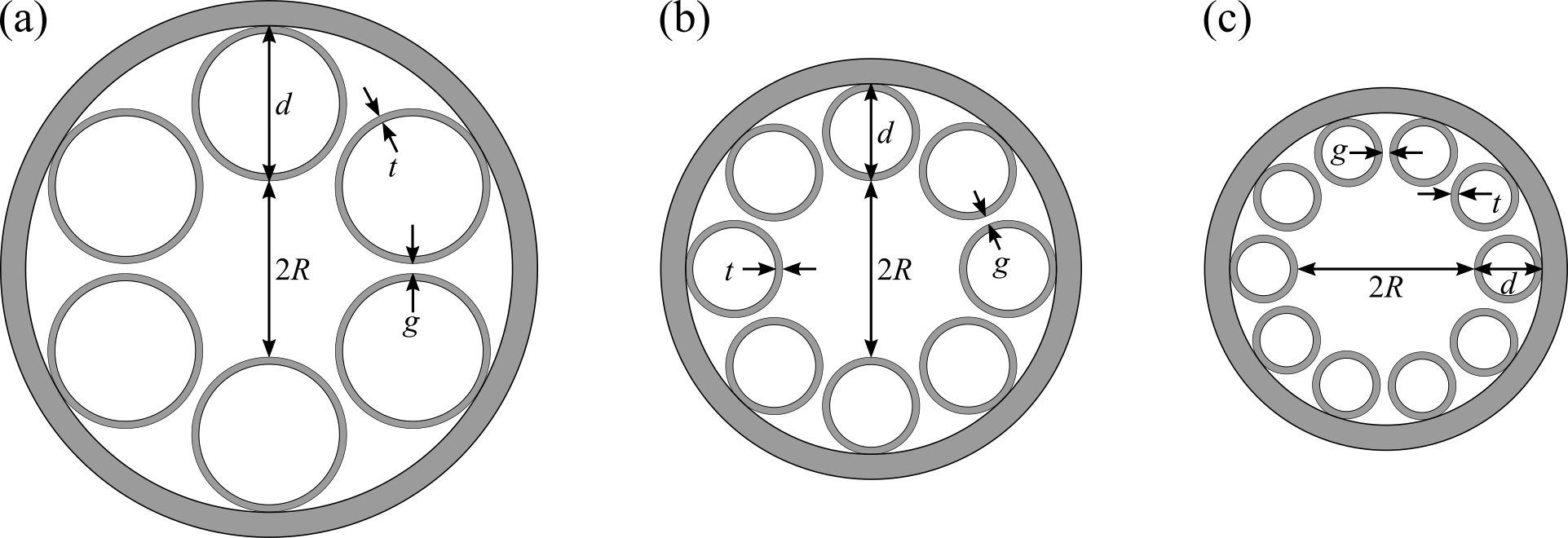}
\caption{Idealized cross-sectional structures of HC-ARFs with (a) $n=6$, (b) $n=8$ and (c) $n=10$. $R$ and $t$ denote the core radius and the glass-web thickness, respectively. The antiresonant tubes have the outer diameter $d$, and $g$ is the perimeter gap.}
\label{fig:fig1}
\end{figure}

Figure \ref{fig:fig1} presents idealized cross sections of nodeless HC-ARFs that are considered in this work. 
Each structure consists of different number of antiresonant tubes $n$, i.e.~(a) $n=6$, (b) $n=8$ and (c) $n=10$. The HC-ARFs have core radius $R$, the glass-web thickness $t$ and the perimeter gap between the antiresonant tubes $g$. Then, the outer diameter of the antiresonant tube $d$ is given by: 
\begin{equation}
d=\frac{2R\sin\left(\pi\mathbin{/}n\right)-g}{1-\sin\left(\pi\mathbin{/}n\right)}\textrm{.}
\label{eq:eq1}
\end{equation}

In order to numerically calculate the guiding properties of these structures, we employ the finite-element method (FEM). For accurate numerical simulations, we use mesh resolution of up to $\lambda/8$ ($\lambda$ is operating wavelength), and apply an optimized perfectly-matched layer on the boundary. By carefully analyzing the numerical results, we were able to obtain empirical formulae for characterizing the guiding properties of a large variety of HC-ARFs over a wide spectral range. Note that in this study, we consider only the fundamental mode.

\section{Dispersion of HC-ARFs}

Our empirical formula for the dispersion is based on the capillary model derived by Marcatili and Schmeltzer \cite{Marcatili1964}. The capillary radius is corrected to account for the HC-ARF's core-cladding structure, as well as the operating wavelength. Moreover, the formula incorporates the structural resonances in accordance with the antiresonant reflecting optical waveguide (ARROW) model, which arise due to the glass-web thickness of the antiresonant tubes. Our formula for the effective index $n_{\textrm{eff}}$ of the HC-ARF's has the form:
\begin{equation}
n_{\textrm{eff}}=\sqrt{1-{\left(\frac{u_{\textrm{01}}\lambda}{2\pi R_{\textrm{eff}}}\right)}^2+\sum\limits_{m}\frac{\sigma_{\textrm{m}}\lambda^2}{\lambda^2-\left(2t\mathbin{/}m\right)}}\textrm{.}
\label{eq:eq2}
\end{equation}
Here, the second term accounts for the effect of the core-cladding arrangement, where $u_{\textrm{01}}\approx2.405$ is the first zero of the Bessel function of the first kind and $R_{\textrm{eff}}$ is the effective radius which we discuss below. The last term denotes the effect of the ARROW model, and accounts for the structural resonances. $m$ is the resonance order and $\sigma_{\textrm{m}}$ describes its corresponding strength. For example, we found that, $\sigma_{\textrm{m}}\approx 10^{\left(-9-m\right)}$, gives a reasonable agreement with the FEM calculations around the resonances. 

The effective radius $R_{\textrm{eff}}$ is given by:
\begin{equation}
R_{\textrm{eff}}=f_{1}R\left(1-\frac{f_{2}\lambda^2}{Rt}\right)\textrm{,}
\label{eq:eq3}
\end{equation}
where $f_1$ and $f_2$ are two dimensionless parameters that need to be determined to provide fitting in the short and long wavelength regions, respectively. Since most HC-ARF applications are carried out at wavelengths far away from the structural resonances, the last term in Eq.~(\ref{eq:eq2}) can be dropped for simplicity. Hence, the formula then becomes:
\begin{equation}
n_{\textrm{eff}}\approx1-\left(\frac{1}{8}\right){\left(\frac{u_{\textrm{01}}\lambda}{\pi R_{\textrm{eff}}}\right)}^2\textrm{.}
\label{eq:eq4}
\end{equation}

\begin{figure}[h] 
\centering
\includegraphics{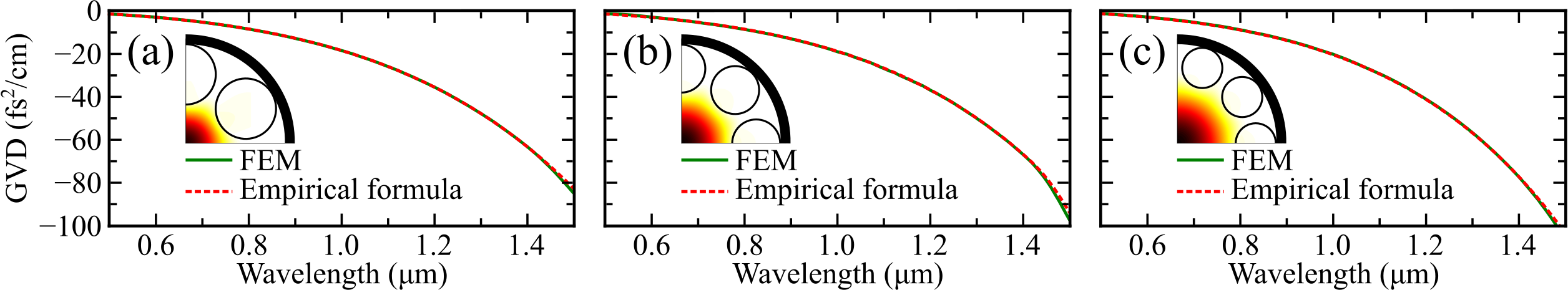}
\caption{The GVD curves calculated using FEM and Eq.~(\ref{eq:eq4}) for HC-ARFs with (a) $n=6$, (b) $n=8$ and (c) $n=10$. Here, $R=\SI{12}{\micro\metre}$, $t=\SI{0.2}{\micro\metre}$ and $g=\SI{1}{\micro\metre}$. For the best fitting $f_1=1.1039$ for all three structures and $f_2=0.0480$, $0.0608$ and $0.0729$ for (a), (b) and (c), respectively.}
\label{fig:fig2}
\end{figure}

Figure \ref{fig:fig2} shows the comparison between the GVD curves calculated numerically using FEM and the empirical formula given in Eq.~(\ref{eq:eq4}), for the three HC-ARF structures with different $n$ given in Fig.~\ref{fig:fig1}.
 The other design parameters are fixed at $R=\SI{12}{\micro\metre}$, $t=\SI{0.20}{\micro\metre}$ and $g=\SI{1}{\micro\metre}$. As for the fitting parameters, $f_1=1.1039$ was used for all three structures, while different $f_2$ values were required for different antiresonant tube arrangements, i.e.~$f_2=0.0480$, $0.0608$, and $0.0729$ for Figs.~\ref{fig:fig2}(a), (b) and (c), respectively. This indicates that the dispersion is not strongly influenced by the cladding arrangement in the short-wavelength side, whereas it becomes important in the longer wavelength.

\begin{figure}[h] 
\centering
\includegraphics{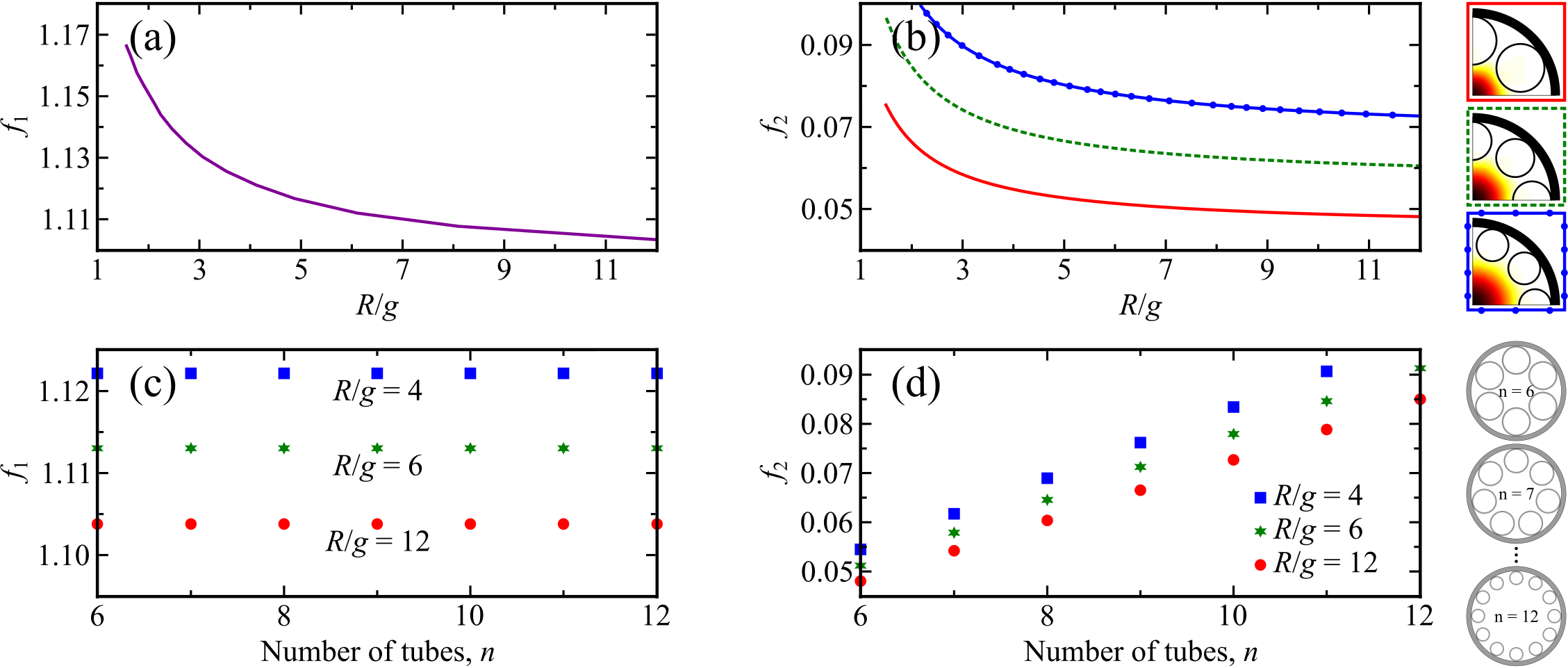}
\caption{The fitting parameters (a) $f_1$ and (b) $f_2$ determined using the best fit as a function of $R\mathbin{/}g$ for HC-ARFs. (c) and (d) present how $f_1$ and $f_2$ changes with $n$.}
\label{fig:fig3}
\end{figure}

In fact, our analysis reveals that these fitting parameters depend on the dimensionless ratio between the core radius and the perimeter gap. Figures \ref{fig:fig3}(a) and (b) present the variations in the fitting parameters $f_1$ and $f_2$ as a function of the ratio $R\mathbin{/}g$ for the HC-ARF with different $n$, shown in Fig.~\ref{fig:fig1}. They clearly show that $f_1$ curve does not vary with the change in $n$, while $f_2$ curve shifts up with the increasing $n$. Figures \ref{fig:fig3}(c) and (d) show how $f_1$ and $f_2$ vary with increasing $n$ from six to twelve for three different $R\mathbin{/}g$ ratios.

From these plots, we find the empirical relationships between the fitting parameters and the ratio $R\mathbin{/}g$, which are given by:
\begin{subequations}
\begin{equation}
f_1=A_1\exp\left(\frac{A_0}{R\mathbin{/}g}\right)\textrm{,}
\label{eq:eq5a}
\end{equation}
\begin{equation}
f_2=B_1n\exp\left(\frac{B_0}{R\mathbin{/}g}\right)-B_2n+B_3\textrm{,}
\label{eq:eq5b}
\end{equation}
\end{subequations}
where $A_0=0.097041$, $A_1=1.095$, $B_0=0.76246$, $B_1=0.007584$, $B_2=0.002$ and $B_3=0.012$.

\begin{figure}[h] 
\centering
\includegraphics{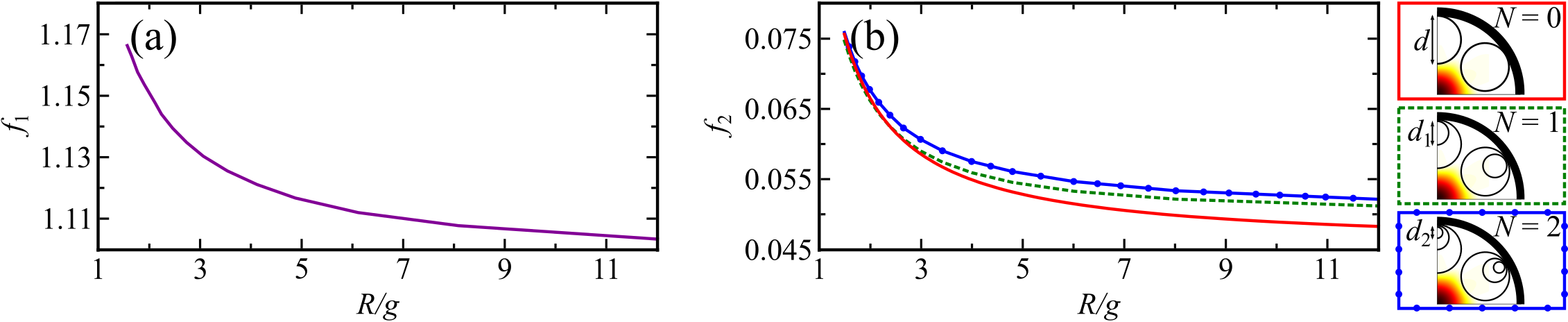}
\caption{The fitting parameters (a) $f_1$ and (b) $f_2$ determined using the best fit as a function of $R\mathbin{/}g$ for HC-ARF with $N=0$. Additional $f_2$ curves are shown in (b) for $N=1$ and $N=2$.}
\label{fig:fig4}
\end{figure}

In past investigations, it was found that having nested elements in the antiresonant tubes improved the guiding properties \cite{Hasan2017}. Hence, we also study the effect of having the nested elements as shown in Fig.~\ref{fig:fig4}. We consider $n=6$ for HC-ARFs with no-nested elements ($N=0$), one-nested elements ($N=1$) and two-nested elements ($N=2$). We set the outer diameters of the nested elements to be $d_1=d\mathbin{/}2$ and $d_2=d_1\mathbin{/}2$, which is known to provide the best light confinement \cite{Belardi2014}.

Similar to Figs.~\ref{fig:fig3}(a) and (b), $f_1$ is not affected by the presence of the nested elements, while $f_2$ is increased when the number of nested elements is increased. This indicates that the dispersion in the short wavelength side is not influenced much by the presence of nested elements, while in the long wavelength region, it becomes more important. Therefore, we can expand Eq.~(\ref{eq:eq5b}) to account for the presence of nested elements, such that it becomes:
\begin{equation}
f_2=B_1n\exp\left(\frac{B_0}{R\mathbin{/}g}\right)-B_2n+B_3+K_{\textrm{f}}\textrm{,}
\label{eq:eq6}
\end{equation}
where $K_{\textrm{f}}$ depends on the number of nested elements $N$, and is given by:
\begin{equation}
K_{\textrm{f}}=0.0045\exp\left(\frac{-4.1589}{N\left(R\mathbin{/}g\right)}\right)\textrm{.}
\label{eq:eq7}
\end{equation}

\section*{Funding}
Australian Research Council (ARC) ( DP140100265 and DP150102057); Volkswagen Stiftung.

\end{document}